\documentclass[aps,prl,preprint]{revtex4}
\pdfoutput=1
\usepackage{graphicx}
\usepackage{amsmath}
\usepackage{multirow}
\usepackage{subfigure}
\usepackage{gensymb}
\usepackage{color}
\usepackage{xcolor}
\usepackage{soul}
\usepackage{braket}
\usepackage{hyperref}
\date{}

\begin{document}
\title { \bf Model-independent test for CPT violation using long-baseline and atmospheric neutrino experiments}
\author{ Daljeet Kaur}
\affiliation {S.G.T.B. Khalsa College, University of Delhi}

\begin{abstract}
Charge-Parity-Time (CPT) symmetry governs that the oscillation parameters for neutrinos and anti-neutrinos are to be identical. Different mass and mixing parameters for these particles may give us a possible hint for CPT violation in the neutrino sector. Using this approach, we discuss the ability of long-baseline and atmospheric neutrino experiments to determine the difference between mass squared splittings ($\Delta m^{2}_{32}-\Delta\bar{m}^{2}_{32}$) and atmospheric mixing angles ($\sin^{2}\theta_{23}-\sin^{2}\bar{\theta}_{23}$) of neutrinos and anti-neutrinos. We show the joint sensitivity of the T2K, NOvA and INO experiments to such CPT violating observables in different possible combinations of octant for neutrinos and anti-neutrinos.
\end{abstract}

\maketitle

\newpage 
\section{Introduction}
The fact that neutrinos have mass and flavour mixed are strongly confirmed with the discovery of neutrino oscillations\cite{SNO1,SK1,SK2,K2K, KAM1}. The existence of neutrino masses is in fact the first solid experimental fact requiring physics beyond the Standard Model. Under the assumption of conservation of the fundamental CPT symmetry, both neutrino and anti neutrino oscillations are described by three mass eigen states $\nu_{1}, \nu_{2}, \nu_{3}$ with mass values $m_{1}$, $m_{2}$ and $m_{3}$ that are connected to the flavor states $\nu_{e}, \nu_{\mu}$ and $\nu_{\tau}$ by a mixing matrix U\cite{bruno,pmns1}. The neutrino or anti-neutrino  oscillation probability depends on three mixing angles, $\theta_{12}$, $\theta_{23}$, $\theta_{13}$; two independent mass differences, $|\Delta m^{2}_{32}|$, $\Delta m^{2}_{21}$; where $\Delta m^{2}_{32}$= $m^{2}_{3}-m^{2}_{2}$ and $\Delta m^{2}_{21}$= $m^{2}_{2}-m^{2}_{1}$; and a CP violating phase $\delta_{CP}$. %Additional phases are present in case neutrinos are Majorana particles, but they do not influence neutrino flavor oscillations at all.
The primary goals of present and future neutrino oscillation experiments are to perform precision measurements of the neutrino parameters, determine the right order of neutrino masses (i.e., the sign of $\Delta m^{2}_{32}$), determine the right octant [Lower Octant (LO) if $\theta_{23}<45^{o}$ and Higher Octant (HO) if $\theta_{23}>45^{o}$] and to determine the value of CP phase $\delta_{CP}$.

With the increasing knowledge of the standard neutrino oscillation parameters, searches for the symmetry-breaking effects become also possible.
For example, with the nonzero value of $\theta_{13}$\cite{Reno,DB}, it became possible to search for CP-violation in the neutrino sector via the differences in the oscillation probabilities of neutrinos and anti-neutrinos. Similarly, CPT violation have been studied by several neutrino oscillation experiments under various assumptions\cite{CPT1,CPT2,CPT3,CPT4,CPT5,CPT6,CPT7,CPT8,CPT9,CPT10,CPT11,CPT12,CPT13,CPT14,CPT15,CPT16,CPT17,CPT18,CPT19,CPT20,CPT21,CPT22}.
According to the conservation of CPT symmetry, the mass-squared splitting and mixing angles are expected to be identical for neutrinos and anti-neutrinos. Therefore, an independent measurement of neutrino and anti-neutrino oscillation parameters and their comparison can be treated as a model independent way to test the CPT-conservation or it could possibly give us a sign for CPT-violation\cite{CPT23,CPT24,CPT25,CPT26,CPT27,CPT28,CPT29}. 

In this paper, we use the model independent way to test the CPT theorem under the standard three neutrino paradigm. We consider the possibility that the oscillation probability  governed by neutrino mass splitting or mixing angle is different  as compared to that of anti-neutrinos. Thus, the differences between neutrino and anti-neutrino oscillation parameters might be regarded as CPT violating observables. We perform realistic simulations for the current and future long-baseline oscillation experiments (T2K, NOvA) and atmospheric neutrino experiment (ICAL-INO). We explore the potential of these experiments to test the CPT conservation and the CPT violation, assuming non-identical neutrino and anti-neutrino oscillation parameters. Since, the octant of neutrinos or anti-neutrinos is still unknown, we also show the potential of these experiments in different possible combinations of octants for neutrinos and anti-neutrinos.

This paper is organized as follows. A brief introduction of the experiments used in the analysis is given in Section~\ref{expt}. In Section~\ref{AM}, we describe the details of simulations work for atmospheric (INO) and long-baseline experiments (T2K and NOvA) separately. In Section~\ref{work}, we show the experimental sensitivity of T2K, NOvA and INO experiments considering CPT is conserved [Subsection~\ref{cpttr}] followed by the CPT violation sensitivities [Subsection~\ref{cptv}]. We explore the joint sensitivity for these experiment under Subsection~\ref{combined}. Finally, we conclude our results in  Section~\ref{results}.

\section{Experimental Specifications}
\label{expt}

\begin{itemize}
\item \textbf {The INO-ICAL Experiment}:
The India-based Neutrino Observatory (INO)\cite{INO3} is an atmospheric neutrino experiment, which will be located at Bodi West hills in the Theni district of South India. A  50 kton magnetised ICAL detector will be the main detector at INO to address the current issues of neutrino physics like neutrino mass hierarchy, octant of $\theta_{23}$ and the precise determination of neutrino mixing parameters. The 1 km rock overburden above the site will act as a natural shield from the background of cosmic rays.  The ICAL detector will be of rectangular shape of dimensions $48m\times16m\times14.5m$ having three modules. Each module weighing about 17 kton with the dimensions $16m\times16m\times14.5m$. Each module will consist of 151 layers of 5.6~cm thick iron plates with alternate gaps of 4 cm where the active detector element will be placed. In the first phase of INO, glass Resistive Plate Chambers (RPCs)\cite{DJ_nim} will be used as active detector to track the charged particles produced through the interaction of muon neutrinos with iron target. Another important feature of the INO-ICAL experiment is the application of a magnetic field of 1.5 T that will help in distinguishing the charge of the interacting particles. This distinction is crucial for the precise determination of relative ordering of neutrino mass states (neutrino mass hierarchy) and other parameters. The INO-ICAL experiment is sensitive to atmospheric muons only. Hence, it will observe interactions of muon type neutrinos. The ICAL experiment will also measure the energy of hadron shower to improve the energy reconstruction of events, and hence the overall sensitivity to neutrino parameters\cite{hdphy,DJ_epjc}.

\item \textbf{The NOvA Experiment}:
The NOvA (NuMi off-axis $\nu_{e}$ appearance)\cite{nova1, nova2} is a long-baseline neutrino experiment that uses an NuMI beam source at Fermilab. It is designed to study the $\nu_{\mu} \rightarrow \nu_{e}$ appearance oscillations and  $\nu_{\mu} \rightarrow \nu_{\mu}$ survival oscillations.  It uses a high intensity proton beam with a beam power of 0.7 MW. It consists of  two detectors; Near Detector (ND) and Far Detector (FD), which  are  functionally  identical  and  14.6  mrad off axis from the Fermilab NuMI beam to receive a narrow-band neutrino energy spectrum near 2 GeV.  The ND is 1 km away from the beam source to detect the unoscillated beam and a 14-kton liquid scintillator FD is located in Ash River, Minnesota, with a baseline of 810 km from the fermilab to detect the oscillated neutrino beam. The long-baseline of NOvA  enhances the matter effect and allows probing of the neutrino mass ordering. The experiment is designed to operate in neutrino mode (using neutrino beam flux) and anti-neutrino mode (using anti-neutrino beam flux). 
     The long base-line oscillation channels used in NOvA includes 1. $\nu_{e}$ appearance, 2. $\nu_{\mu}$ disappearance, 3. NC disappearance. NOvA has the potential to measure the precise value of neutrino mixing angles, determine neutrino mass hierarchy  and  can  investigate  the  CP  violation  in  the  lepton sector. It is scheduled to run 5 years in $\nu$ mode followed by 5 years in $\bar\nu$ mode.

 \item \textbf{The T2K Experiment}: 
 The T2K (Tokai to Kamioka) \cite{T2K1, T2K2} experiment is a long-baseline neutrino oscillation experiment.  The experiment uses an intense proton beam of 0.77 MW power generated by the J-PARC accelerator in Tokai, Japan. T2K composed of a neutrino beamline, a near detector complex (ND280), and a far detector (Super-Kamiokande) located 295 km away from J-PARC.
 T2K is an off-axis experiment which generate the narrow-band neutrino beam using proton synchrotron at J-PARC. The off-axis angle is set at 2.5 degree so that the narrow-band $\nu_{\mu}$ beam peaks at energy of 2 GeV, which maximizes the effect of the neutrino oscillation at 295 km and minimizes the background to electron neutrino appearance detection. 
 The near detector site at nearly 280 m from the production target and houses on-axis and off-axis detectors. The on-axis detector (INGRID), composed of an array of iron/scintillator sandwiches, measures the neutrino beam direction and profile. %The off-axis detector, immersed in a magnetic field, measures the muon neutrino flux and energy spectrum, and intrinsic electron neutrino contamination in the beam in the direction of the far detector, along with measuring rates for exclusive neutrino reactions. These measurements are essential in order to characterize signals and backgrounds that are observed in the Super-Kamiokande far detector.
 The off-axis detector is composed of a water-scintillator detector, the tracker consisting of time projection chambers (TPCs) and fine grained detectors (FGDs) optimized to study charged current interactions; and an electromagnetic calorimeter (ECal). The whole off-axis detector is placed in a 0.2 T magnetic field.
 The far detector, Super-Kamiokande, is located at Kamioka Mine, Japan. The detector cavity lies under the peak of a mountain, with 1000 m of rock overburden. It has a 22.5 kt water Cherenkov detector consisting of a welded stainless steel tank, 39 m in diameter and 42 m tall. The detector contains approximately 13,000 photomultiplier tubes (PMTs) that image neutrino interactions in pure water. The main goal of T2K experiment is to measure the last unknown lepton sector mixing angle $\theta_{13}$ by observing $\nu_{e}$ appearance in a $\nu_{\mu}$ beam. It also aims to make a precision measurement of the known oscillation parameters, $|\Delta m^{2}_{32}|$ and $\theta_{23}$, via $\nu_{\mu}$ disappearance studies. Other goals of the experiment include various neutrino cross-section measurements and sterile neutrino searches.
\end{itemize}

\section{Analysis Methodology}
\label{AM}
\begin {itemize}
  \item \textbf{For atmospheric neutrino experiment}:
The magnetized ICAL detector enables separation of neutrino and anti-neutrino interactions for atmospheric events, allowing an independent measurement of the $\nu_{\mu}$ and $\bar{\nu}_{\mu}$
oscillation parameters. We analyze the reach of the Iron Calorimeter for $\nu_{\mu}$ and $\bar{\nu}_{\mu}$ oscillations separately using a three flavor analysis including the Earth 
matter effects. A large number of unoscillated NUANCE\cite{nuance} neutrino events have been generated using HONDA\cite{honda} atmospheric neutrino fluxes for an exposure of 50 kt $\times$ 1000 years of the ICAL detector. Analysis has been performed by normalizing these events to 500 kt-yr exposure for the ICAL detector. Each Charged-Current (CC) neutrino event is characterized by its energy and zenith angle. Oscillation effects have been introduced via a Monte-Carlo reweighting algorithm as described in earlier works\cite{DJ_epjc,trk,moonmoon}.

Each oscillated neutrino or anti-neutrino event is divided as a function of twenty muon energy bins ($E_{\mu}$), twenty muon zenith angle ($\cos\theta_{\mu}$) and five hadron 
energy bins ($E_{hadron}$) of optimized bin width as mentioned in Ref.\cite{prd}. These binned data are then folded with detector efficiencies and resolution functions as provided by the INO collaboration\cite{mureso,hreso} for the reconstruction of neutrino and anti-neutrino events separately. We use a ``pulled'' $\chi^{2}$\cite{maltoni} method based on Poisson probability distribution to compare the expected and observed data. The functions $\chi^2(\nu_{\mu})$ and $\chi^2(\overline{\nu}_{\mu})$ are calculated separately for the independent measurement of neutrino and anti-neutrino oscillation parameters. The two $\chi^{2}$ can be added to get the
%Each $\chi^2$ is fitted with 20 muon energy bins, 20 muon angle bins and 5 hadron energy bins via $20\times20\times5 =2000$ binning scheme for neutrino as well as for anti-neutrinos.  
combined $\chi^2(\nu_{\mu}+\overline{\nu}_{\mu})$ as
 \begin{equation}
\label{eq:chiino}
  \chi^2(\nu_{\mu}+\overline{\nu}_{\mu}) =\chi^{2}(\nu_{\mu}) + \chi^{2}(\overline{\nu}_{\mu}).
 \end{equation}

The $\nu$ and $\overline{\nu}$ events are separately binned into direction and energy bins. For different energy and direction bins, the $\chi^{2}$ function is minimized with respect to these four parameters along with the nuisance parameters to take the systematic uncertainties into account as considered in earlier ICAL analyses\cite{trk, DJ_epjc}. Other simulation inputs are summerised as shown in Table~\ref{tb_INO}.

\begin{table}[]
  \centering
  \renewcommand\arraystretch{0.5}
\resizebox{\textwidth}{!}{%
\begin{tabular}{|l|l|}
\hline
\textbf{Characteristics}                                                          & \textbf{INO}                                                                                                                                                                               \\ \hline
Source                                                                            & Atmospheric Neutrinos                                                                                                                                                                      \\ \hline
Run time                                                                          & 10 years for $\nu_{\mu}$ and $\bar{\nu}_{\mu}$                                                                                                                                             \\ \hline
Detector                                                                          & 50kton Iron Calorimeter                                                                                                                                                                    \\ \hline
\begin{tabular}[c]{@{}l@{}}Charge identification efficiency\end{tabular}        & $\sim99\%$ for $\mu^{-}$ and $\mu^{+}$ for few GeV muons as given in Ref.\cite{mureso}                                                                                                                             \\ \hline
\begin{tabular}[c]{@{}l@{}}Direction reconstruction efficiency\end{tabular} & $\sim1^{\degree}$ for few GeV muons as in Ref\cite{mureso}                                                                                                                                              \\ \hline
Systematics                                                                       & \begin{tabular}[c]{@{}l@{}}$20\%$ flux normalisation, $10\%$ cross-section, $5\%$ tilt error,\\ $5\%$ zenith angle error and $5\%$ overall systematics error  as in Refs.\cite{trk, DJ_epjc}\end{tabular} \\ \hline
\end{tabular}%
}
\caption{Experimental specifications used in the analysis for INO atmospheric neutrino experiment.}
\label{tb_INO}
\end{table}

\item \textbf{For Long-baseline neutrino experiments}:
The beamline experiments are suitable for both neutrino and anti-neutrino mode, it is easy to study the sensitivity for the oscillation parameters for neutrino and anti-neutrino independently. In order to quantify the sensitivities of  the long-baseline experiments T2K and NOvA experimental setups, we use GLoBES\cite{globes1, globes2} as a simulator.
For the NovA experiment simulations, we use 3 years $\nu$ and 3 years $\bar\nu$ running mode with beam power of 0.7MW with 20e20 POT/year.
The NOvA detector properties considered in this analysis are taken as in Ref.~\cite{sanjeeba}.
We have considered input files for T2K from the General Long Baseline Experiment Simulator (GLoBES) package\cite{globes1,globes2} and the updated experimental description of T2K are taken from\cite{t2k_gl1,t2k_gl2}. In this analysis, we have used 5 years $\nu$ and 5 years $\bar\nu$ running modes for T2K with beam power of 0.75MW.
    We analyse the  neutrino events from $\nu_{e}$ appearance and $\nu_{\mu}$ disappearance oscillation channels and anti-neutrino events from $\bar{\nu_{e}}$ appearance and $\bar{\nu_{\mu}}$ disappearance oscillation channels. For parameter-estimation, we make use of a chi-squared statistics that is a function of independent physics parameters for neutrinos and anti-neutrinos. For a given set of neutrino and anti-neutrino oscillation parameters, we compute the expected number of signal and background events as a function of energy for the experiment of interest. The values of $\chi^{2}$ are evaluated for $\nu$ and $\bar{\nu}$ separately using the standard rules as described in GLoBES. Other detailed description of simulation inputs are shown in Table~\ref{Tb_inputs}.

  \begin{table}[htbp]
    \centering
         \renewcommand\arraystretch{0.5}
\resizebox{\textwidth}{!}{%
\begin{tabular}{|l|l|l|}
\hline
\textbf{Characteristics} & \textbf{NOvA}                                                                                                                                                                                     & \textbf{T2K}                                                                                                                                                                                    \\ \hline
Baseline                 & 810km                                                                                                                                                                                             & 295km                                                                                                                                                                                           \\ \hline
Run time                 & 3 year $\nu$ and 3 year $\bar{\nu}$                                                                                                                                                               & 5 year $\nu$ and 5year $\bar{\nu}$                                                                                                                                                              \\ \hline
Detector                 & 14 kton                                                                                                                                                                                           & 22.5 kton                                                                                                                                                                                       \\ \hline

signal efficiency        & \begin{tabular}[c]{@{}l@{}}26\% for $\nu_{e}$ and 41\% $\bar{\nu_{e}}$ signal\\ 100\% for both $\nu_{\mu}$ CC and $\bar{\nu_{\mu}}$ CC\end{tabular}                                                              & \begin{tabular}[c]{@{}l@{}}87\% for both $\nu_{e}$ and $\bar{\nu_{e}}$ signal\\ 100\% for both $\nu_{\mu}$ CC and $\bar{\nu}_{\mu} $CC\end{tabular}                                                            \\ \hline
Background efficiency    & \begin{tabular}[c]{@{}l@{}}0.83\%$\nu_{\mu}$ CC, 0.22\%$\bar{\nu_{\mu}}$CC\\ 2\%$\nu_{\mu}$ NC, 3\% $\bar{\nu_{\mu}}$NC\\ 26\%(18\%)$\nu_{e}$ and $\bar{\nu_{e}}$ beam contamination\end{tabular} & considered as given in Refs.~\cite{t2k_gl1, t2k_gl2} \\ \hline
Systematics              & \begin{tabular}[c]{@{}l@{}}5\% signal normalization error\\ 10\% background normalization error\end{tabular}                                                                                      & \begin{tabular}[c]{@{}l@{}}2\% signal normalization error\\ 20\% background normalization error\end{tabular}                                                                                    \\ \hline
\end{tabular}%
}
\caption{Experimental specifications used in the analysis for Long-Baseline experiments.}
\label{Tb_inputs}
\end{table}

\end{itemize}

\section{Analysis}
\label{work}
%We perform our analysis in following steps:

%(1)\textbf{Test for the CPT symmetry:} Assuming identical oscillation parameters for neutrino and anti-neutrinos, we show the experimental capabilities to measure the oscillation parameters for neutrinos and anti-neutrinos independently.[This will consider as null hypothesis for our analysis]\\
%(2)\textbf{Test for the CPT violation:} Assuming non-identical oscillation parameters for neutrinos and anti-neutrinos we show the experimental capabilities to rule out the null hypothesis by measuring the differences between neutrinos and anti-neutrinos oscillation parameters.

\subsection {Neutrino and anti-neutrino oscillation parameters}
Here, we introduce the notation used to describe neutrino and anti-neutrino oscillations used in the analysis. We use the neutrino oscillation parameters as three mixing angles, $\theta_{12}$, $\theta_{23}$, $\theta_{13}$; two independent mass differences, $\Delta m^{2}_{32}$, $\Delta m^{2}_{21}$, and a CP phase $\delta_{CP}$. Similarly, anti-neutrino parameters are described with a bar over them as three mixing angles, $\bar{\theta}_{12}$, $\bar{\theta}_{23}$, $\bar{\theta}_{13}$; two independent mass differences, $\Delta \bar{m}^{2}_{32}$, $\Delta \bar{m}^{2}_{21}$, and a CP phase $\bar{\delta}_{CP}$. The analysis considers only normal mass ordering, therefore only positive values of $\Delta m^{2}_{32}$ or $\Delta \bar{m}^{2}_{32}$ have been used. For discussing differences between neutrino and anti-neutrino oscillation parameters, we use notation $\Delta(x)=x-\bar{x}$; where $x$ is any oscillation parameters. So, $\Delta{x}=0$ corresponds to identical oscillation parameters for $\nu$ and $\bar{\nu}$ or CPT conserved assumption and $\Delta{x} \neq 0$ corresponds to the CPT violation assumption. Since all the experiments considered in this paper are quite sensitive to atmospheric oscillation parameters so  we mainly discuss the experimental sensitivities for finding out the difference between the atmospheric mass squared splittings i.e. $\Delta(\Delta m^{2}_{32})$=$\Delta m^{2}_{32}$-$\Delta\bar{m}^{2}_{32}$ and mixing angle difference i.e. $\Delta(\sin^{2}\theta_{23})$= $ \sin^{2}\theta_{23}$-$\sin^{2}\bar{\theta}_{23}$.

The global best fit values of oscillation parameters which are kept fixed through out the analysis are given as : $\sin^2\theta_{13} (\bar{\theta}_{13})$=0.0234, $\sin^2\theta_{12}(\bar\theta_{12})$=0.313,
$\Delta m^{2}_{12}(\Delta\bar m^{2}_{12})$=7.6 $\times$ $10^{-5}$ eV$^{2}$. Since, the ICAL is insensitive to the variation of $\delta_{CP}$ phase\cite{cpsense}, hence it is kept fixed at $0^{\degree}$. However, NOvA and T2K are sensitive to $\delta_{CP}$ so we marginalized the $\delta_{CP}$ in range 0-$360^{\degree}$ for the predicted data set. To find the sensitivities for atmospheric mass-squared splittings and mixing angles, oscillation parameters ($\Delta m^{2}_{32}, \Delta\bar{m}^{2}_{32}, \sin^{2}\theta_{23}$ and $\sin^{2}\bar{\theta}_{23}$) are allowed to fit in the range given in Table~\ref{osc_tb2}.

\begin{table}[htbp]
\begin{center}
\begin {tabular}{c c}
\hline
\hline
Oscillation parameters &  allowed fit range  \\
\hline 
\hline
$\Delta m^{2}_{32}$ (eV$^{2}$) & (2.0-3.0) $\times$ $10^{-3}$\\
$\Delta\bar{m}^{2}_{32}$ (eV$^{2}$) & (2.0-3.0) $\times$ $10^{-3}$  \\
$ \sin^2\theta_{23}$ & 0.3-0.7 \\
$ \sin^2\overline {\theta}_{23}$ &   0.3-0.7 \\
\hline 
\end {tabular}
\caption{\label{osc_tb2}The neutrino and anti-neutrino oscillation parameters and their range.} 
\end{center}

\end{table}

\subsection{Test for the CPT symmetry}
\label{cpttr}
In this section, we discuss the capabilities of NOvA, T2K and INO experiments to test the CPT-theorem. We show how well the neutrino and anti-neutrino can be measured independently from one another assuming CPT is a good symmetry. We consider the oscillation parameters for $\nu$ and $\bar{\nu}$ are identical and show the allowed regions for the parameters of interest assuming CPT is conserved. This identical parameters ($\nu -\bar{\nu} =0$) is then taken as null hypothesis for analysis presented in Section~\ref{cptv}.\\
%In order to explore the experimental capabilities to test the CPT-theorem,
We test the sensitivities for  $\nu$ oscillation parameters ($\Delta m^{2}_{32}, \sin^{2}\theta_{23}$) and  $\bar{\nu}$ oscillation parameters $( \Delta\bar{m}^{2}_{32}, \sin^{2}\overline{\theta}_{23}$). In order to do so we proceed as follows.
First, a fake dataset is generated at the fixed true values of $\nu$ or $\bar{\nu}$ oscillation parameters and then a two dimensional grid search is performed for the predicted dataset in the allowed ranges of the parameters as mentioned in Table~\ref{osc_tb2}. Further, $\chi^2$ is calculated between the fake dataset and predicted dataset for each set of true values of oscillation parameters. The functions $\chi^2(\nu)$ and $\chi^2(\overline{\nu})$ are calculated separately as an independent measurement of $\nu$ and $\bar{\nu}$. A joint result from the combined neutrino and anti-neutrino analysis is also shown. The two $\chi^{2}$ can be added to get the combined analysis results $\chi^2(\nu+\bar{\nu})$ as mentioned in equation~\ref{eq:chiino}.

\begin{figure}[htbp]
  
% \centering
 \subfigure[]{
  \includegraphics[width=0.30\textwidth,height=5cm]{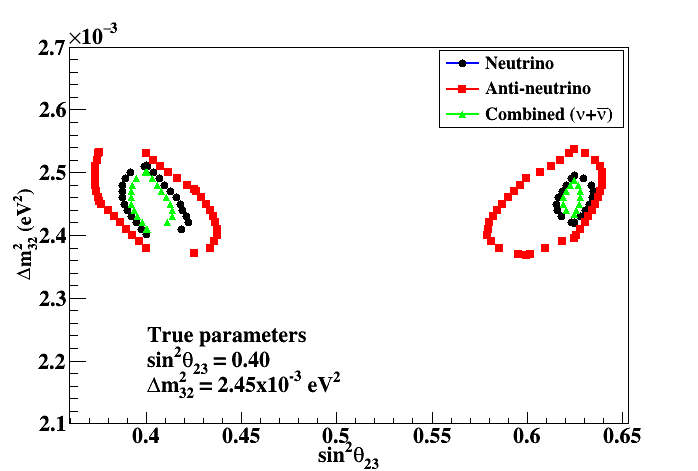}
   \label{fig:a}
   }
 \subfigure[]{
  \includegraphics[width=0.30\textwidth,height=5cm]{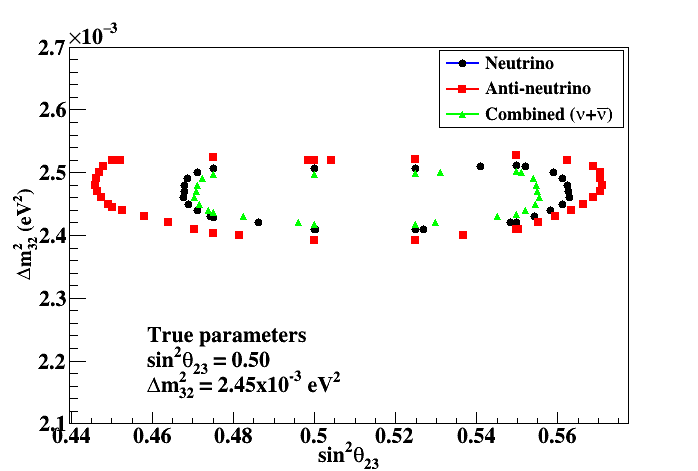}
   \label{fig:b}
   }
\subfigure[]{
  \includegraphics[width=0.30\textwidth,height=5cm]{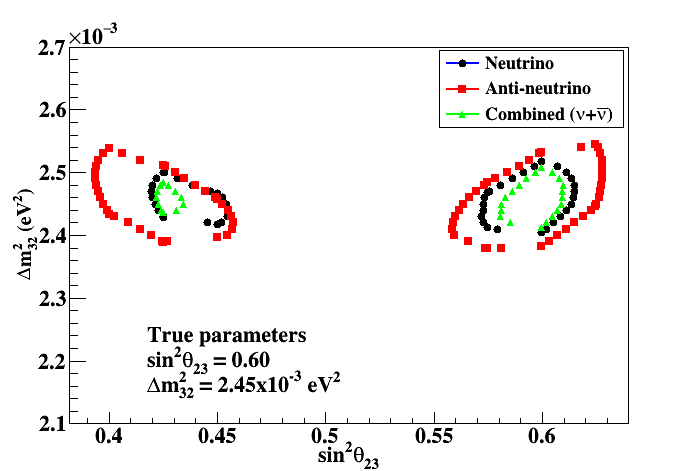}
   \label{fig:c}
}
\caption{\label{fig1}90$\%$  C.L. expected region obtained from NOvA experiment for lower octant ($sin^{2} \theta_{23}=0.40$)[Left], maximal mixing ($sin^{2} \theta_{23}=0.50$)[Middle], and for higher octant ($sin^{2} \theta_{23}=0.60$)[Right] with $\Delta m^{2}_{32}=2.45 \times10^{-3} eV^{2}$, asuming CPT is conserved. Red, black and green contours are obtained as a results of anti-neutrino, neutrino and combined ($\nu+\bar\nu$) analysis respectively.}
\end{figure}

\begin{figure}[htbp]
  
% \centering

\subfigure[]{
  \includegraphics[width=0.30\textwidth,height=5cm]{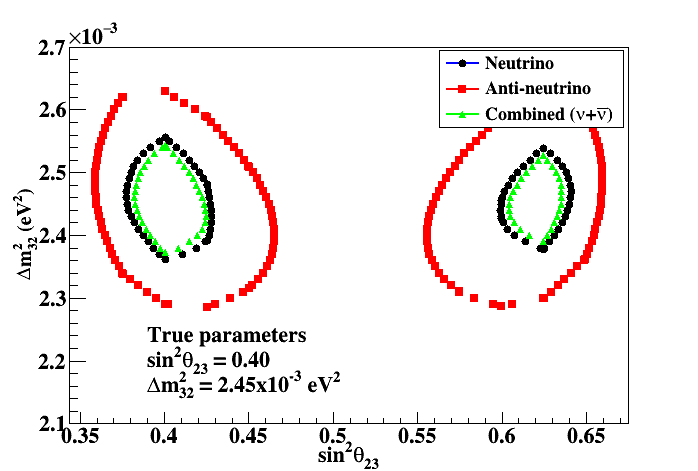}
   \label{fig:d}
}
\subfigure[]{
  \includegraphics[width=0.30\textwidth,height=5cm]{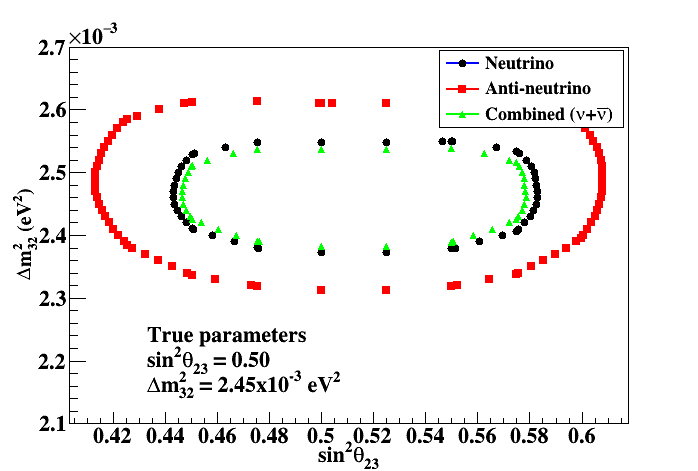}
   \label{fig:e}
}
\subfigure[]{
  \includegraphics[width=0.30\textwidth,height=5cm]{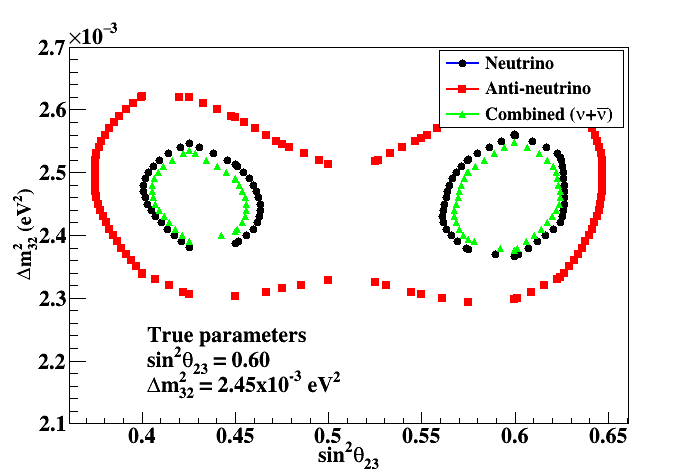}
   \label{fig:f}
   }
\caption{\label{fig2}90$\%$  C.L. expected region obtained from T2K experiment for lower octant ($sin^{2} \theta_{23}=0.40$)[Left], maximal mixing ($sin^{2} \theta_{23}=0.50$)[Middle],  and for higher octant ($sin^{2} \theta_{23}=0.60$)[Right] with $\Delta m^{2}_{32}=2.45 \times10^{-3} eV^{2}$, asuming CPT is conserved. Red, black and green contours are obtained as a results of anti-neutrino, neutrino and combined ($\nu+\bar\nu$) analysis respectively.}
\end{figure}

\begin{figure}[htbp]
   \subfigure[]{
  \includegraphics[width=0.30\textwidth,height=5cm]{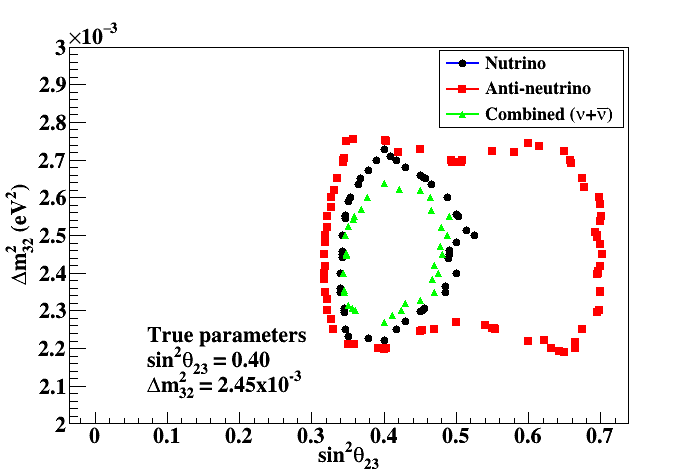}
   \label{fig:aa}
   }
 \subfigure[]{
  \includegraphics[width=0.30\textwidth,height=5cm]{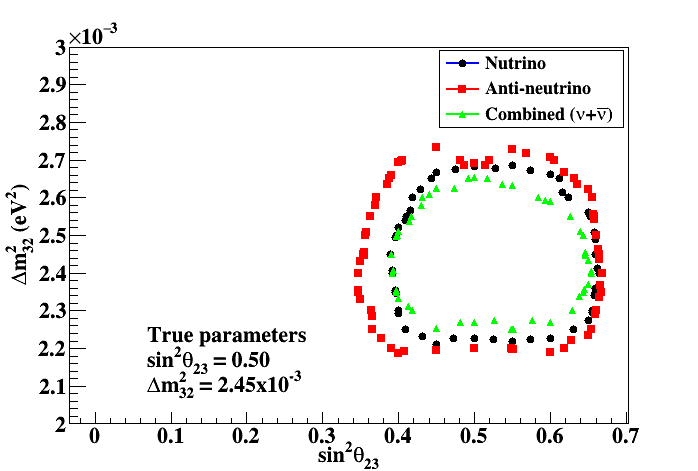}
   \label{fig:bb}
   }
\subfigure[]{
  \includegraphics[width=0.30\textwidth,height=5cm]{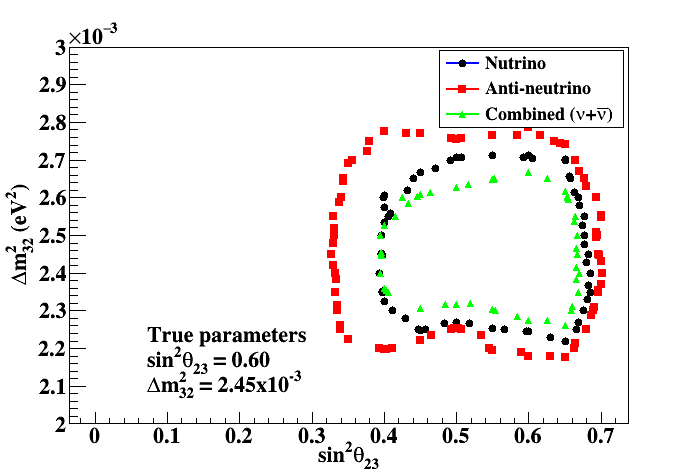}
   \label{fig:cc}
}
\caption{\label{fig3}90$\%$  C.L. expected region obtained from INO experiment for lower octant ($sin^{2} \theta_{23}=0.40$)[Left], maximal mixing ($sin^{2} \theta_{23}=0.50$)[Middle],  and for higher octant ($sin^{2} \theta_{23}=0.60$)[Right] with $\Delta m^{2}_{32}=2.45 \times10^{-3} eV^{2}$, asuming CPT is conserved. Red, black and green contours are obtained as a results of anti-neutrino, neutrino and combined ($\nu+\bar\nu$) analysis respectively.}
\end{figure}

\begin{table}[htbp]
  \centering
   \renewcommand\arraystretch{0.5}
\resizebox{0.65 \textwidth}{!}{%
\begin{tabular}{|c||c|c|c||c|c|c|}
  \hline
  \multicolumn{1}{|c||}{Analysis Mode} & \multicolumn{3}{|c||}{$\Delta m^{2}_{32} (or \Delta \bar{m}^{2}_{32})$ in \%} & \multicolumn{3}{|c|}{$\sin^{2}{\theta_{23}}({or \bar{\theta}_{23}})$ in \%} \\ \hline
 Experiments & NOvA & T2K & INO & NOvA & T2K & INO \\ \hline
Anti-Neutrinos & 2.43 & 6.15 & 11.02 & 11.50  & 19.00  & 30.61  \\ \hline
Neutrinos  & 1.95  & 3.61 & 9.11 & 8.83 & 13.65 & 25.97 \\ \hline
Combined ($\nu+\bar{\nu}$) & 1.56 & 3.19 & 7.80 & 7.97 & 12.90 & 25.26 \\ \hline
\end{tabular}
}
\caption{Precision measurement of parameters $\Delta m^{2}_{32} (\Delta \bar{m}^{2}_{32})$  and $\sin^{2}{\theta_{23}}({\bar{\theta}_{23}})$ for NOvA, T2K and INO experiment for maximal mixing $\sin^{2}{\theta_{23}}({\bar{\theta}_{23}})=0.5$ and $\Delta m^{2}_{32} (\Delta \bar{m}^{2}_{32})=2.45 \times 10^{-3} eV^{2}$.}
\label{tb:prec_tb}
\end{table}

%We show the samples of the simulated neutrinos and anti-neutrino events with true values as: $\sin^2\theta _{23}({\bar\theta}_{23}$ =0.4,0.5 and 0.6 known as Lower octant (LO), maximal mixing(MM) and Higher octant(HO) respectively with $\Delta m^{2}_{32}(|\Delta \bar{m}^{2}_{32}|$=2.45 $\times 10^{-3}eV^2$.

The results of the neutrino, anti-neutrino  and  their joint data analyses have been shown on a single frame  projecting over two-dimensional regions with allowed regions at $90\%$ Confidence Level (CL)  in the atmospheric plane ($\Delta m^{2}_{32}(\bar{m^{2}}_{32}), \sin^{2}\theta_{23}(\bar{\theta}_{23})$).

Figure~\ref{fig1}, \ref{fig2} and \ref{fig3} show the expected sensitivities obtained from NOvA, T2K and INO experiments respectively having best fit values as $\sin^2\theta _{23}({\bar\theta}_{23})$ = 0.4 [Lower Octant (LO), 0.5 [Maximal Mixing (MM)] and 0.6 [Higher Octant (HO)] with $\Delta m^{2}_{32} (\Delta \bar{m}^{2}_{32})=2.45 \times 10^{-3} eV^{2}$.
  
  Results are shown for LO, MM and for HO as left, middle and right plots respectively. It can be  observed from these sample plots that for all the mentioned experiments, there is a clear difference between neutrino's and anti-neutrino's parameters space when they are analyzed independently. Neutrino only analysis give more stringent or precise parameter's space comparable to anti-neutrino only analysis. However, the $\nu+\bar{\nu}$ joint results are found be more precise as compare to independent $\nu$ and $\bar{\nu}$ analyses. An overall comparison of the precision for the measurement of ($\Delta m^{2}_{32}, \sin^{2}\theta_{23}$) obtained from these experiments at the maximal mixing is shown in Table~\ref{tb:prec_tb}. We would like to mention that precision measurement of these parameters is not the main focus of this paper but it is interesting that assuming CPT is conserved, there is a difference between  the independent measurement of neutrino and anti-neutrino oscillation parameters as shown in Table~\ref{tb:prec_tb}. This motivates us to do the CPT violation test where we can use this study as our null hypothesis. 
  
The octant of $\nu$ and $\bar{\nu}$ plays an important role in the neutrino and anti-neutrino parameter estimation. One can observe a clear octant degeneracy from Figure~\ref{fig1} and Figure~\ref{fig2}. The NOvA experiment clearly shows two degenerate solutions of $\sin^2\theta_{23}$ at the lower octant as well as  higher octant [Figure~\ref{fig:a}, Figure~\ref{fig:c}] in all the analyses (neutrino, anti-neutrino and combined ($\nu$+$\bar{\nu}$)). However, T2K experiment shows a clear octant degeneracy at the LO in all the analyses [Figure~\ref{fig:d}], while at the higher octant side, two degenrate solutions exist only in anti-neutrino and combined ($\nu$+$\bar{\nu}$) analyses [Figure~\ref{fig:f}]. Figure~\ref{fig3} depicts that INO does not show any octant degeneracy in the mixing angle.

%\clearpage

\subsection{Test for the CPT violation}
\label{cptv}
 We study the NOvA, T2K and INO experiment's sensitivity to measure CPT violation by  determining how well these experiments can rule out the conserved CPT assumption for neutrino and anti-neutrino parameters.
For this, we started with the  assumption that neutrino and anti-neutrinos have different mass-squared splittings and mixing angles such that the difference
$[\Delta (\Delta {m}^{2}_{32})=(\Delta m^{2}_{32}-\Delta \bar{m^{2}}_{32}) \neq 0]$, and $[\Delta \sin^{2}\theta_{23} = (\sin^2\theta_{23}-\sin^2\bar\theta_{23}) \neq 0] $. To rule out the null hypothesis i.e. identical oscillation parameters for neutrinos and anti-neutrinos, a fake dataset is generated at a given set of true values of neutrino and anti-neutrino oscillation parameters ($\Delta m^{2}_{32}$, $ \sin^2\theta_{23}$,  $\Delta \bar{m}^{2}_{32}$, $ \sin^2\bar\theta_{23}$). A four dimensional grid search  is performed for the predicted dataset. $\chi^2$ is calculated between the fake dataset and predicted dataset for each set of true values of oscillation parameters. Now, the true values of the oscillation parameters are not fixed at single value rather it also varied  in the range as  mentioned in Table~\ref{osc_tb2} and same procedure is repeated again for each set of true values. We calculated $\Delta (\Delta {m}^{2}_{32})$ and $\Delta \sin^{2}\theta_{23}$. To find out the sensitivity for the difference $\Delta (\Delta {m}^{2}_{32})$, a minimum $\chi^2$ has been binned as a function of difference in the true values of $\Delta(\Delta {m}^{2}_{32})$ keeping marginalization over $\Delta \sin^{2}\theta_{23}$ and for the sensitivity for difference of mixing angles $\Delta \sin^{2}\theta_{23}$, same has been done with the marginalization over $\Delta(\Delta{m}^{2}_{32})$. Further, for each set of difference $\Delta (\Delta {m}^{2}_{32})$ or $\Delta \sin^{2}\theta_{23}$, we calculate $\Delta\chi^{2}=\chi^{2}-\chi^{2}_{min}$ and plot it as the functions of set of differences.
 
It is quite possible that in nature neutrino and anti-neutrino may lie in same or different octant. We also try to simulate the data considering this possibility to obtained the detector sensitivity for $\Delta (\Delta {m}^{2}_{32})$ and $\Delta \sin^{2}\theta_{23}$   in combination of different octants. There are four possible combinations of octants for neutrino and anti-neutrinos:  
\\Case 1: $\nu$s and $\bar{\nu}$s both in Higher Octant (HO) [$\sin^{2}{\theta_{23}}(\sin^{2}{\bar{\theta}_{23}})$ in range 0.5-0.7]\\
Case 2: $\nu$s and $\bar{\nu}$s both in Lower Octant (LO) [$\sin^{2}{\theta_{23}}(\sin^{2}{\bar{\theta}_{23}})$ in range 0.3-0.5] \\
Case 3: $\nu$s in HO and $\bar{\nu}$s in LO \\
Case 4: $\nu$s in LO and $\bar{\nu}$s in HO

\begin{figure}[htbp]
 \centering
 \subfigure[]{
   \includegraphics[width=0.48\textwidth,height=7cm]{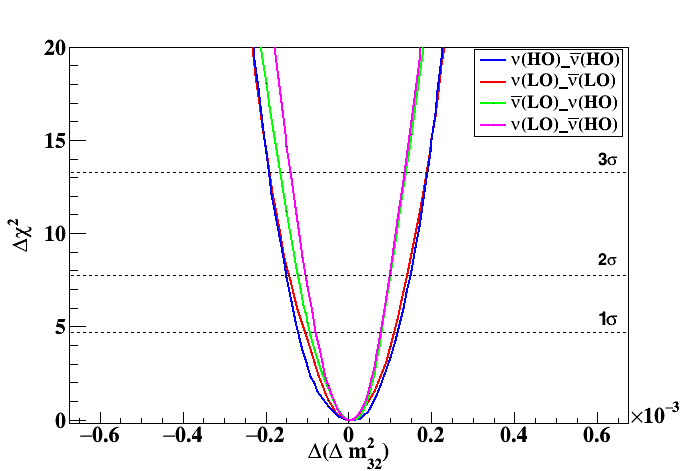}
   \label{fig:g} 
   }
 \subfigure[]{
  \includegraphics[width=0.48\textwidth,height=7cm]{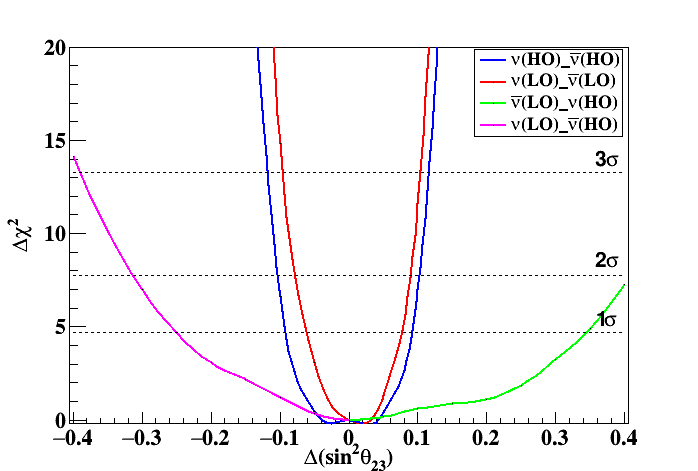}
   \label{fig:h}
 }
\caption{\label{fig:Nova_1d}NOvA experiment sensitivity for  $\Delta(\Delta m^{2}_{32})eV^{2}$ [Left]  and  $\Delta \sin^{2}\theta_{23}$ [Right] for different possible combinations of octant for $\nu$ and $\bar\nu$ having non-identical oscillation parameters.  }
\end{figure}

\begin{figure}[htbp]
 \centering
\subfigure[]{
  \includegraphics[width=0.48\textwidth,height=7cm]{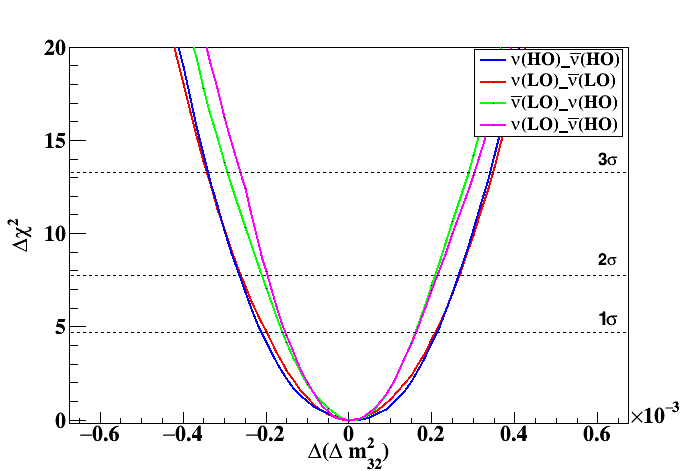}
   \label{fig:i}
   }
\subfigure[]{
  \includegraphics[width=0.48\textwidth,height=7cm]{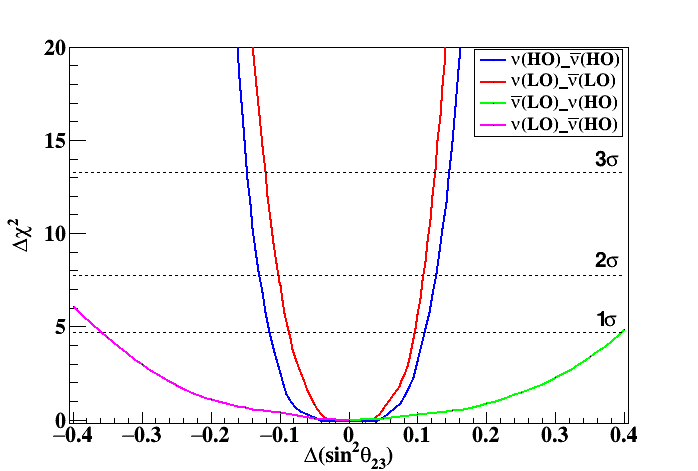}
   \label{fig:j}
   }
\caption{\label{fig:T2k_1d}The T2K experiment sensitivity for  $\Delta(\Delta m^{2}_{32})eV^{2}$ [Left]  and  $\Delta \sin^{2}\theta_{23}$ [Right] for different possible combinations of octant for $\nu$ and $\bar\nu$ having non-identical oscillation parameters.}
 \end{figure}

\begin{figure}[htbp]
 \centering
\subfigure[]{
  \includegraphics[width=0.48\textwidth,height=7cm]{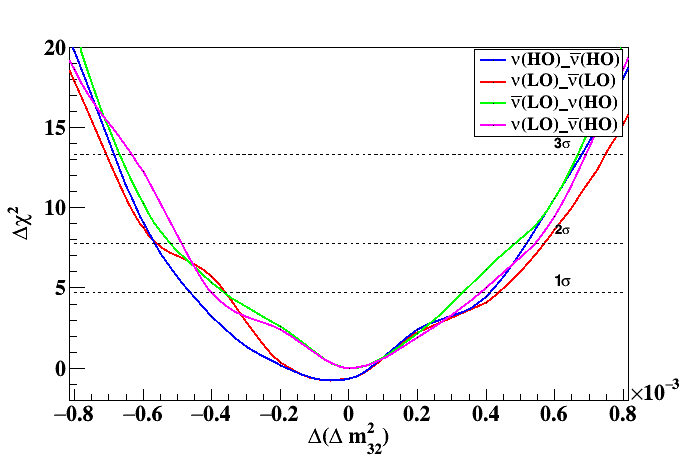}
   \label{fig:k}
   }
\subfigure[]{
  \includegraphics[width=0.48\textwidth,height=7cm]{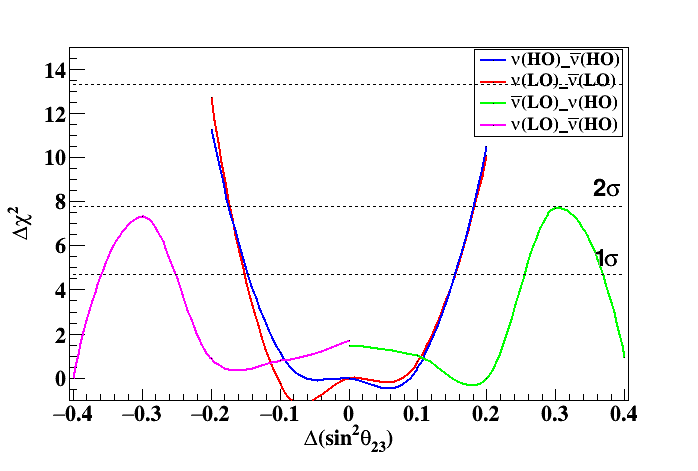}
   \label{fig:l}
   }
\caption{\label{fig:INO_1d}The INO experiment sensitivity for  $\Delta(\Delta m^{2}_{32})eV^{2}$ [Left]  and  $\Delta \sin^{2}\theta_{23}$ [Right] for different possible combinations of octant for $\nu$ and $\bar\nu$ having non-identical oscillation parameters. }
 \end{figure}

Figure~\ref{fig:Nova_1d},\ref{fig:T2k_1d} and \ref{fig:INO_1d} show the one dimensional experimental sensitivities of  $\Delta(\Delta{m}^{2}_{32})$  and  $\Delta \sin^{2}\theta_{23}$ for the NOvA, T2K and INO experiments respectively.

Figure~\ref{fig:Nova_1d}[Left] shows the NOvA sensitivity for the mass squared splitting difference parameter $\Delta(\Delta m^{2}_{32})$ for all possible cases of octants for $\nu$ and $\bar{\nu}$ as mentioned earlier. It has been observed that for case 3 and case 4 (where $\nu$ and $\bar{\nu}$ are assumed to be in different octant) gives slightly better sensitivity for $\Delta(\Delta m^{2}_{32})$  than the similar octant combinations (case1 and case2).  NOvA can rule out the CPT conserved scenario by measuring  $\Delta(\Delta m^{2}_{32})$ with 2$\sigma$ significance level $\sim 0.15 \times 10^{-3} eV^{2}$ for the similar octant combinations (case 1 and case2) and it is  $\sim 0.10 \times 10^{-3} eV^{2}$ for different octant combination (case 3 and case4).

The right panel of Figure~\ref{fig:Nova_1d} shows the NOvA sensitivity for $\Delta \sin^{2}\theta_{23}$. It is found that the NOvA is most sensitive for $\Delta \sin^{2}\theta_{23}$ only if the neutrinos and anti-neutrinos are in same octant (either LO or HO) and out of this, case 2, where  $\nu$ and $\bar{\nu}$ both in Lower octant (LO) gives the slightly better sensitivity ($>3\sigma$ when $|\Delta \sin^{2}\theta_{23}|=0.08$]) than case 1 where $\nu$ and $\bar{\nu}$ both in Higher octant (HO). But, if neutrino and anti-neutrino octants are different, the sensitivity is almost $<2\sigma$ in the range [-0.4, 0] of $\Delta \sin^{2}\theta_{23}$ for octant case 3 [$\nu$ in HO and $\bar{\nu}$ in LO] and it is $<3\sigma$ in the range [0, 0.4)] of $\Delta \sin^{2}\theta_{23}$ for octant case 4 [$\nu$ in LO and $\bar{\nu}$ in HO]. So we can say that for the NOvA experiment, similar octants for $\nu$ and $\bar{\nu}$ are favourable.

 Figure~\ref{fig:T2k_1d} shows the sensitivities to $\Delta(\Delta m^{2}_{32})$  and  $\Delta \sin^{2}\theta_{23}$ for the T2K experiment. The left panel of Figure~\ref{fig:T2k_1d} depicts the T2K sensitivity to rule out the CPT conserved scenario [$\Delta(\Delta m^{2}_{32})$] for all possible combinations of octants assumed for neutrinos and anti-neutrinos. We observed that for T2K, opposite octant combination (case 3 and case4)  for $\nu$ and $\bar{\nu}$ gives slightly better sensitivity than the similar octant combinations (case 1 and case 2).
T2K can rule out the CPT conserved scenario by measuring  $\Delta(\Delta m^{2})$ as $0.2 \times 10^{-3} eV^{2}$ with 2$\sigma$ significance level for the similar octant combinations (case 1 and case2) and it is $\sim 0.275 \times 10^{-3} eV^{2}$ for different octant combination (case 3 and case4).

The right panel of Figure~\ref{fig:T2k_1d} shows that T2K sensitivity for the difference of mixing angles $\Delta \sin^{2}\theta_{23}$.
 Similar to the NOvA experiment, the T2K experiment is also found to be most sensitive for the ($\Delta \sin^{2}\theta_{23}$) only if $\nu$s and $\bar{\nu}$s are in same octant (either LO or HO) and  case 2 gives the better results than case 1. If  different octant assumed for neutrino and anti-neutrinos [case 3 and case 4], T2K sensitivity for $\Delta \sin^{2}\theta_{23}$ is almost $<1\sigma$ in the given range. So we can say that similar to the NOvA experiment, case 1 and case 2 are also most favourable for T2K experiment.

Similarly, Figure~\ref{fig:INO_1d} show the sensitivities to $\Delta(\Delta m^{2}_{32})$  and  $\Delta \sin^{2}\theta_{23}$ for the atmospheric INO-ICAL experiment. It is clear for the left panel of the Figure~\ref{fig:INO_1d} that ICAL detector can rule out the null hypothesis of identical mass-squared splittings for neutrino and anti-neutrinos  with 2$\sigma$ significance level for almost all possible combinations of octants if the $|\Delta(\Delta m^{2}_{32})|$ is roughly around $0.5 \times 10^{-3} eV^{2}$. And, similar to the NOvA and T2K experiment, it is also found to be least sensitive for the difference of neutrino and anti-neutrino mixing angles ($\Delta \sin^{2}\theta_{23}$) for the octant case 3 and 4 [Figure~\ref{fig:l}]. For similar octant combinations for neutrino and anti-neutrino, the sensitivity of the ICAL detector is almost similar to NOvA and T2K experiment and can rule out the identical mixing angles for neutrino and anti-neutrino with 2$\sigma$ significance level if the $|\Delta \sin^{2}\theta_{23}|=0.08$.

 \subsection{Combined Experimental Sensitivities for $\Delta(\Delta m^{2}_{32})$ and $\Delta \sin^{2}\theta_{23}$ }
 \label{combined}

\begin{figure}[htbp]
 \centering
  \includegraphics[width=0.5\textwidth,height=7cm]{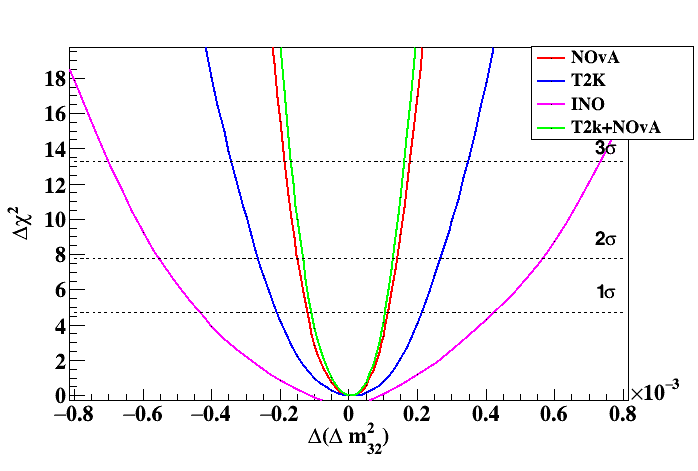}
   \caption{\label{fig:m}Experimental sensitivity of the NOvA, T2K and the INO  experiments for $\Delta(\Delta m^{2}_{32})eV^{2}$.}
  \end{figure}

 As it is clear from Section~\ref{cptv}, with the considered exposure and run time, the NOvA and T2K experiment's sensitivity  is quite better compared to the INO-ICAL experimental sensitivity. Hence, we also show a combined long base-line (T2K and NOvA) sensitivity for a better estimation of $\Delta(\Delta m^{2}_{32})$ and $\Delta \sin^{2}\theta_{23}$. We observed that $\Delta(\Delta m^{2}_{32})$is not affected from different octant considerations for neutrinos and anti-neutrinos. So, we show an overall estimation for the measurement of $\Delta(\Delta m^{2}_{32})$ [Figure~\ref{fig:m}] from the NOvA, T2K and INO-ICAL experiments. 

\begin{figure}[htbp]
 \centering
\subfigure[case 1]{
  \includegraphics[width=0.48\textwidth,height=7cm]{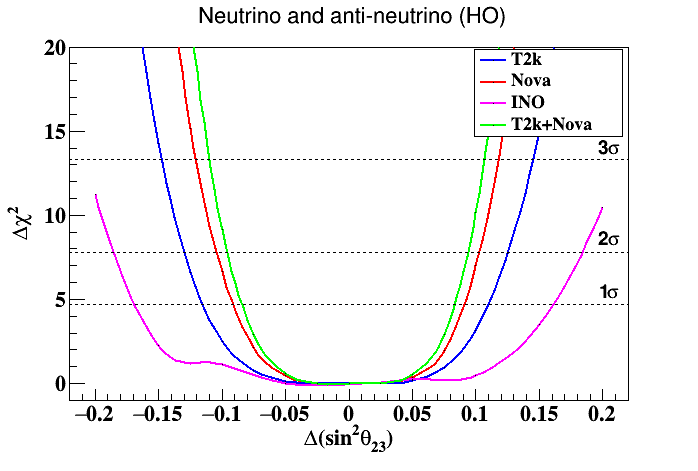}
  \label{fig:p}
}
\subfigure[case 2]{
  \includegraphics[width=0.48\textwidth,height=7cm]{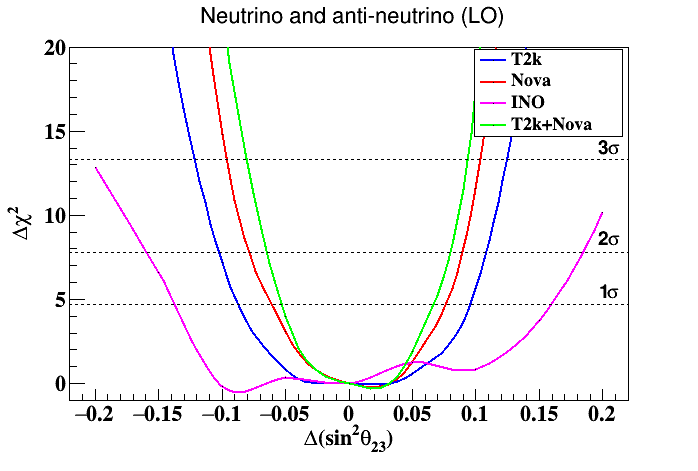}
   \label{fig:o}
   }
\subfigure[case 3]{
  \includegraphics[width=0.48\textwidth,height=7cm]{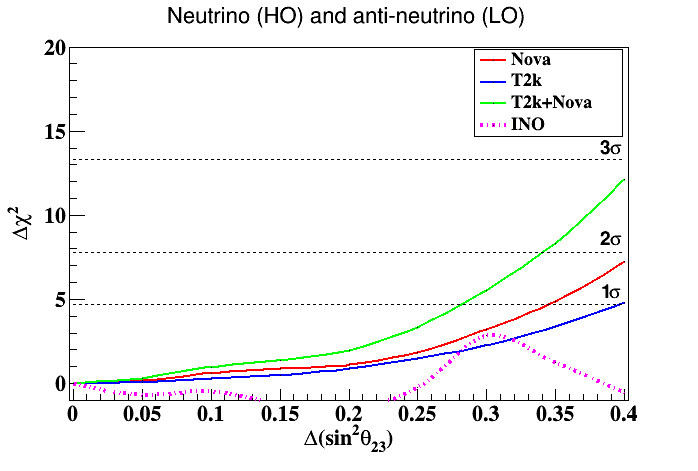}
  \label{fig:q}
}
\subfigure[case 4]{
  \includegraphics[width=0.48\textwidth,height=7cm]{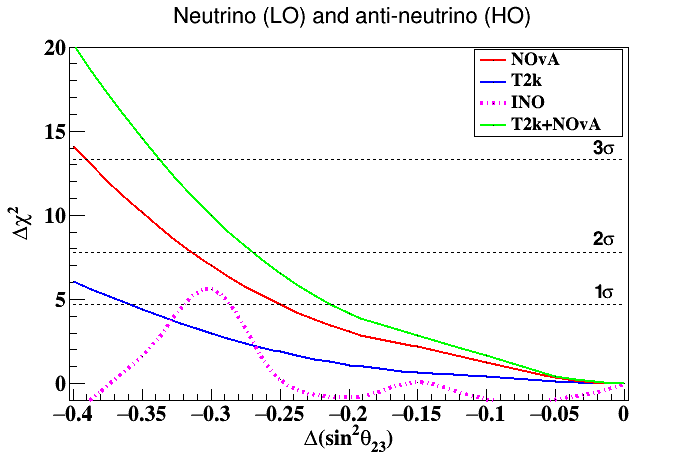}
   \label{fig:n}
   }
\caption{\label{sample1}Combined sensitivity of the NOvA, T2K and INO  experiments for $\Delta \sin^{2}\theta_{23}$ =$sin^{2} \theta_{23}-sin^{2} \bar{\theta_{23}}$ when  (a) $\nu$ and $\bar{\nu}$ in HO, (b) $\nu$ and $\bar{\nu}$ in LO, (c) $\nu$ in HO and $\bar{\nu}$ in LO and (d) when $\nu$ in LO and $\bar{\nu}$ in HO.  }
\end{figure}

  A quantitative comparison of potential of these experiments for $\Delta(\Delta m^{2}_{32})$ is shown in Table~\ref{tb:th}. It is clear from Figure~\ref{fig:m} and Table~\ref{tb:th} that the NOvA sensitivity is almost comparable to joint (NOvA+T2K) sensitivity for  $\Delta(\Delta m^{2}_{32})$. We expect that NOvA experiment itself can able to rule out the identical oscillation parameters (CPT is conserved) by measuring  $\Delta(\Delta m^{2}_{32})$ in comparison to NOvA+T2K combined analyses. 
  Similarly, Figure~\ref{sample1} shows the combined sensitivity of the NOvA, T2K and INO  experiments for the measurement of  $\Delta \sin^{2}\theta_{23}$ in different possible combination of octant as mentioned in section~\ref{cptv}. Here, we find that although NOvA experiment is good enough to constraine $\Delta \sin^{2}\theta_{23}$ for case 1 and case 2, but on combining T2K and NOvA data, the sensitivity for $\Delta \sin^{2}\theta_{23}$ significantly increases for octant case 3 and case 4, where neutrinos and anti-neutrinos are assumed to be in different octant. A quantitative comparison of the sensitivity for $\Delta \sin^{2}\theta_{23}$ in different octants is shown in Table~\ref{tb:th}.

\begin{table}[]
  \centering
  \renewcommand\arraystretch{0.5}
  \resizebox{0.5\textwidth}{!}{%
    \begin{tabular}{|c|c|c|c|c|}
      \hline

      \multicolumn{5}{|c|}{ $|\Delta(\Delta m^{2}_{32})| \times 10 ^{-3} eV^{2}$ } \\ \hline
      Osc.parameter    & NOvA    & T2K    & INO    & T2K+NOvA   \\ \hline
      $|\Delta(\Delta m^{2}_{32})|$ & 0.10           & 0.22           & 0.40           & 0.10  \\ \hline
      \multicolumn{5}{|c|}{$|\Delta \sin^{2}\theta_{23}|$ } \\ \hline
   Octant Case 1    & 0.1     & 0.13   & 0.16   & 0.07       \\
   Octant Case 2     & 0.08    & 0.12   & 0.17   & 0.09       \\
   Octant Case 3     & 0.34   & 0.4   & $\textless 1\sigma$     & 0.28       \\
   Octant Case 4     & 0.24    & 0.36    &  $\textless 1\sigma$    & 0.21  \\ \hline

\end{tabular}%
}
\caption{$|\Delta(\Delta m^{2}_{32})|$ and $|\Delta \sin^{2}\theta_{23}|$ sensitivity at the 1$\sigma$ confidence level.}
\label{tb:th}
\end{table}

  \section{Summary and Conclusions}
  \label{results}
In this paper we have performed a comprehensive comparative analysis for the CPT violation sensitivities using long-baseline (NOvA and T2K) and atmospheric neutrino (the INO-ICAL) experiments.
  First, we explored how well neutrino and anti-neutrino oscillation parameters are independently constrained by these experiments. Further, we estimated the potential of these experiments to test the hypothesis that neutrino and anti-neutrino oscillation parameters are identical, as governed by the CPT theorem.  We presented a detailed discussion on the sensitivities for the CPT violation observables  ($\Delta(\Delta m^{2}_{32})$ and $\Delta \sin^{2}\theta_{23}$) assuming four possible cases of octants for neutrinos and anti-neutrinos. We show that the experiments (NOvA, T2K and INO-ICAL) are able to constrained these observables for all possible combinations of octants. Individually each experiment is able to measure $\Delta(\Delta m^{2}_{32})$ quite significantly irrespective of different octant combinations, but the measurement of $\Delta \sin^{2}\theta_{23}$ is largely affected by the existence of neutrinos and anti-neutrinos in particular octant. We observed that all considered experiments are giving precise determination of $\Delta \sin^{2}\theta_{23}$ if both neutrinos and anti-neutrinos are assumed to have similar octant combinations (either LO or HO) and these experiments are least sensitive for different octant combinations for neutrinos and anti-neutrinos.
  So, we can say that similar octant combination (either LO or HO) for $\nu$ and $\bar{\nu}$ is favourable condition for precise determination of $\Delta \sin^{2}\theta_{23}$ for all considered experiments.
   One can get a better sensitivity for the estimation of $\Delta(\Delta m^{2}_{32})$ and $\Delta \sin^{2}\theta_{23}$  significantly if we combine the results from different experiments. We study the joint sensitivity of both the long-baseline experiments (T2K+NOvA). Our study shows that with the proposed fiducial volume and run time, the NOvA detector independently found the best among  all the considered experiments for constraining these parameters as shown in Table~\ref{tb:th}. NOvA sensitivity is almost comparable to joint (NOvA+T2K) sensitivity for $\Delta(\Delta m^{2}_{32})$. However, NOvA+T2k joint results enhances the sensitivities for $\Delta \sin^{2}\theta_{23}$ if the neutrinos and anti-neutrinos are in different octants. The present CPT bounds at 1$\sigma$ confidence interval are summarized in Table~\ref{tb:th}.

\section{Acknowledgements}
The author would like to thank University of Delhi R$\&$D grants for providing the support for this research. Author would also like to thank  Dr. Sanjeev Kumar, Dr. Md. Naimuddin, Prof. Brajesh Chandra Choudhary and Prabhjot Singh for many fruitful discussions related to this work.

\end{document}